\documentclass[aps,prc,twocolumn,showpacs]{revtex4-1}

\usepackage{graphicx}
\usepackage{dcolumn}
\usepackage{bm}
\usepackage{ulem}
\usepackage{color}

\newcommand{\al}{$\alpha$}
\newcommand{\g}{$\gamma$}

\newcommand{\rgn}{($\gamma$,n)}

\newcommand{\rpg}{(p,$\gamma$)}

\newcommand{\zrii}{$^{92}$Zr}
\newcommand{\moii}{$^{92}$Mo}
\newcommand{\nbi}{$^{91}$Nb}
\newcommand{\nbii}{$^{92}$Nb}
\newcommand{\nbiii}{$^{93}$Nb}

\newcommand{\spro}{$s$-process}
\newcommand{\rpro}{$r$-process}
\newcommand{\rppro}{$rp$-process}
\newcommand{\ppro}{$p$-process}

\begin{document}

\title{
  Nucleosynthesis of $^{92}$Nb and the relevance of the low-lying isomer at
  135.5\,keV 
}

\author{Peter Mohr}
\email[Email: ]{WidmaierMohr@t-online.de; mohr@atomki.mta.hu}
\affiliation{
Diakonie-Klinikum, D-74523 Schw\"abisch Hall, Germany}
\affiliation{
Institute for Nuclear Research (ATOMKI), H-4001 Debrecen, Hungary}

\date{\today}

\begin{abstract}
\begin{description}
\item[Background] Because of its half-life of about 35 million years,
  $^{92}$Nb is considered as a chronometer for nucleosynthesis events prior to
  the birth of our sun. The abundance of $^{92}$Nb in the early solar system
  can be derived from meteoritic data. It has to be compared to theoretical
  estimates for the production of $^{92}$Nb to determine the time between the
  last nucleosynthesis event before the formation of the early solar system.
\item[Purpose] The influence of a low-lying short-lived isomer on the
  nucleosynthesis of $^{92}$Nb is analyzed. The thermal coupling between the
  ground state and the isomer via so-called intermediate states affects the
  production and  survival of $^{92}$Nb.
\item[Method] The properties of the lowest intermediate state in $^{92}$Nb
  are known from experiment. From the lifetime of the intermediate state and
  from its decay branchings, the transition rate from the ground state to the
  isomer and the effective half-life of $^{92}$Nb are calculated as a function
  of the temperature.
\item[Results] The coupling between the ground state and the isomer is
  strong. This leads to thermalization of ground state and isomer in the
  nucleosynthesis of $^{92}$Nb in any explosive production scenario and almost
  100\,\% survival of $^{92}$Nb in its ground state. However, the strong
  coupling leads to a temperature-dependent effective half-life of $^{92}$Nb
  which makes the $^{92}$Nb survival very sensitive to temperatures
  as low as about 8\,keV, thus turning $^{92}$Nb at least partly into a
  thermometer.
\item[Conclusions] The low-lying isomer in $^{92}$Nb does not affect the
  production of $^{92}$Nb in explosive scenarios. In retrospect this validates
  all previous studies where the isomer was not taken into account. However,
  the dramatic reduction of the effective half-life at temperatures below
  10\,keV may affect the survival of $^{92}$Nb after its synthesis in
  supernovae which are the most likely astrophysical site for the
  nucleosynthesis of $^{92}$Nb.
\end{description}
\end{abstract}

\maketitle

\section{Introduction}
\label{sec:intro}
\nbii\ is a slightly neutron-deficient odd-odd ($Z = 41$, $N = 51$) nucleus
with a long half-life of $t_{1/2} = 3.47 \times 10^7$\,y. It decays
preferentially by electron capture to $^{92}$Zr whereas the energetically also
possible $\beta^-$-decay to $^{92}$Mo was not observed. The long half-life of
\nbii\ results from the high $J^\pi = 7^+$ of the \nbii\ ground state which
suppresses $\beta$-decays to low-lying states in \zrii\ or \moii . Contrary to
the \nbii\ ground state, the first excited state of \nbii\ at $E^\ast =
135.5$\,keV has a low $J^\pi = 2^+$. Thus, direct electromagnetic
M5 or E6 transitions to the
$J^\pi = 7^+$ ground state are highly suppressed, and the $2^+$ isomer decays
also by electron capture to \zrii\ with a short half-life of $t_{1/2} =
10.15$\,d which is 9 orders of magnitude smaller than the half-life of the
\nbii\ ground state. Except explicitly noted, all properties of \nbii\ have
been taken from the ENSDF database \cite{ENSDF} which is based on the Nuclear
Data Sheets \cite{Bag12,Bag00,Bag92,Luk80}.

Unstable nuclei with half-lives of the order of several ten million years are
considered as potential chronometers for the time between the last
nucleosynthesis event and the birth of our sun. For this purpose the
production of these unstable nuclei is compared to the abundance in the early
solar system which can be derived from meteoritic data (e.g.,
\cite{Lug16,Lug14,Dau11,Scho02,Yin00,Har96,Mins82}). Unfortunately, in the
case of \nbii , the production remains relatively uncertain. It is clear that
\nbii\ cannot be produced in the two main neutron capture processes. The
nucleosynthesis path in the slow neutron capture process (\spro ) bypasses
\nbii , and \nbii\ is shielded from production in the rapid neutron capture
process (\rpro ) by the stable isobar \zrii . The stable isobar \moii\ shields
\nbii\ also from production under conditions of the rapid proton capture
process (\rppro ). Thus, only very few astrophysical processes remain as
candidates for the nucleosynthesis of \nbii\ which are in particular the
so-called \ppro\ and neutrino-induced nucleosynthesis. 

Under \ppro\ conditions
\nbii\ may be produced by the \nbiii \rgn \nbii\ reaction although further
destruction by the \nbii \rgn \nbi\ reaction could occur. Alternatively, the
$^{91}$Zr\rpg \nbii\ reaction may produce \nbii\ in a proton-rich
\ppro\ environment. As astrophysical sites for the \ppro , supernova (SN)
explosions of core-collapse type (SN type-II) or thermonuclear explosions of
white dwarfs (SN type Ia) have been suggested (e.g.,
\cite{Rau13,Arn03,Tra14,Dau03}). Very recently it has been emphasized that
\nbii\ can be used to constrain models of \ppro\ nucleosynthesis
\cite{Pig16}. 

Neutrino-induced nucleosynthesis of
\nbii\ occurs via the \zrii ($\nu_e$,$e^-$)\nbii\ or \nbiii ($\nu$,$\nu'
n$)\nbii\ reactions where the high neutrino flux is provided by a forming
neutron star after a core-collapse SN \cite{Sie16,Hay13,Che12,Arc11,Mey03}. In
any case, the nucleosynthesis of \nbii\ occurs in an explosive scenario with
timescales of the order of 1 second and temperatures of at least 1 billion
Kelvin ($T_9 > 1$, $kT \approx 86$\,keV).

The nucleosynthesis of \nbii\ may be significantly influenced by the
properties of the low-lying $J^\pi = 2^+$ isomer at $E^\ast = 135.5$\,keV
which is coupled to the $J^\pi = 7^+$ ground state via so-called intermediate
states (IMS). Interestingly, most of the previous studies did not take into
account this isomer; up to now only Meyer \cite{Mey03} has pointed out that
``Accurate predictions of the \nbiii /\nbii\ production ratio, then, will
require an appropriate treatment of the isomeric state in the nucleosynthesis
calculations''. In general, three questions have to be answered: ($i$) What is
the production ratio between $7^+$ ground state and $2^+$ isomer? ($ii$) How
is the production ratio during the explosive production affected by thermal
couplings via the IMS? ($iii$) Does the coupling via the IMS affect the later
survival of \nbii ? In advance, I provide the answers to these questions from
the following discussion of the properties of the IMS with $J^\pi = 4^+$ at
$E^\ast = 480.3$\,keV: ($i$) and ($ii$): The coupling between $7^+$ ground
state and $2^+$ isomer is so strong that thermal equilibrium between ground
state and isomer is reached almost instantaneously and maintained at least
down to temperatures of the order of 20\,keV. These low temperatures
correspond to an almost negligible Boltzmann ratio of $n(2^+)/n(7^+) = (5/15)
\times \exp{(-E^\ast/kT)} \lesssim 10^{-4}$. Thus, almost 100\,\% of the
produced \nbii\ survives, independent of the production ratio of isomer and
ground state. ($iii$) The strong coupling between $2^+$ isomer and $7^+$
ground state leads to a strongly temperature-dependent effective half-life of
\nbii . Already at $kT = 8.3$\,keV, the effective half-life is reduced by one
order of magnitude, and at $kT = 10.4$\,keV the reduction of the effective
half-life reaches a factor of 1000. This may turn \nbii\ from a chronometer into
a sensitive thermometer for the thermal history between the nucleosynthesis of
\nbii\ in an explosive astrophysical event and the formation of the early
solar system.

\section{Coupling between the ground state and the isomer via intermediate
  states}
\label{sec:coup}
\begin{figure}[htb]
\includegraphics[width=0.80\columnwidth,clip=]{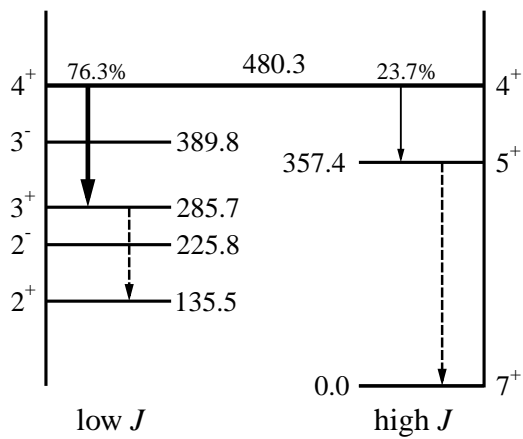}
\caption{
\label{fig:level}
Partial level scheme of \nbii\ (approximately to scale; from
\cite{ENSDF}). The IMS with $J^\pi = 4^+$ at $E^\ast = 480.3$\,keV is marked
by a bold horizontal line. Primary decays from the IMS are shown with full
arrows; secondary \g -rays are shown with dashed arrows. Except for the ground
state, the spin asignments are only tentative but ``very probably'' according
to a comment in \cite{ENSDF}. Therefore, the parentheses in the tentative
$(J)^\pi$ assignments are omitted in this work. The influence of $J^\pi$
assignments on the results of this work is restricted to statistical weights
and thus remains limited when compared to the dominating Boltzmann factors.
}
\end{figure}

Because of the spin difference $\Delta J = 5$ between the $7^+$ ground state
and $2^+$ isomer, a direct \g -ray (M5 or E6) transition is strongly
suppressed by the electromagnetic selection rules. Any transition between
ground state and isomer must proceed via IMSs at higher excitation energies
which are excited by thermal photons. The number of photons in a thermal
stellar photon bath decreases exponentially with energy $E$; thus, typically
the IMS with the lowest excitation energy dominates the transition rates at
stellar temperatures. A careful inspection of the level scheme of \nbii\ shows
that the lowest IMS is the $4^+$ state at $E^\ast = 480.3$\,keV. All levels
below the IMS with $J < 4$ decay finally to the $2^+$ isomer (``low-$J$
states'' in Fig.~\ref{fig:level}, left part) whereas the only level with $J > 
4$ below the IMS decays only to the $7^+$ ground state (``high-$J$ states'').

It is interesting to note that there are only two further states in
\nbii\ above the IMS and below an excitation energy of 1\,MeV. The $6^+$ state
at 501.3\,keV is a high-$J$ state which decays exclusively to the $7^+$ ground
state, and the $(1^+, 2^-)$ state at 975.0\,keV is a low-$J$ state which
decays exclusively to the $2^-$ state at 225.8\,keV. Thus, the $4^+$ state at
480.3\,keV is the only IMS in \nbii\ below $E^\ast = 1$\,MeV, and the coupling
between ground state and isomer under stellar conditions is essentially
defined by this IMS.

The lifetime and the decay branches of the $4^+$ IMS are known from
experiment \cite{ENSDF}. The decay branches have been measured in several
independent ($p$,$n$\g ), ($d$,$t$\g ), and (\al ,$n$\g ) experiments, and good
agreement was found \cite{ENSDF}: $b_\gamma = 23.7 \pm 3.2$\,\% for the
transition to the $5^+$ state at $E^\ast = 357.4$\,keV and $b_\gamma = 76.3 \pm
3.2$\,\% for the transition to the $3^+$ state at $E^\ast = 285.7$\,keV. The
$5^+$ state decays with 100\,\% to the $7^+$ ground state whereas the $3^+$
state decays with 100\,\% to the $2^+$ isomer (see Fig.~\ref{fig:level}).

The lifetime of the IMS is reported only in an unpublished laboratory
report USIP-74-17 of the University of Stockholm which is fortunately
available on the web \cite{USIP}. A half-life of $t_{1/2} = 0.62 \pm 10$
(without units) is given in Table VI of \cite{USIP}. However, from the text in
\cite{USIP} it becomes obvious that the half-life is 0.62\,ps, corresponding
to M1 transition strengths of the order of several $\mu_N^2$ for the above
mentioned transitions. Surprisingly, a half-life of 0.62\,ns has been adopted
in \cite{Luk80}, and this value persists until now
\cite{ENSDF,Bag12,Bag00,Bag92}.

At stellar temperatures, thermal equilibrium within states with low $J$ on the
one hand and within high $J$ states on the other hand is established almost
instantaneously because the levels are connected by allowed \g
-transitions. As a consequence, the transition rate from the $7^+$ ground
state to the $2^+$ isomer depends on the stellar integrated cross section
$I_\sigma^\ast$ of the IMS
\begin{equation}
I_\sigma^{\ast} = g \, \left(\frac{\pi \hbar c}{E^\ast}\right)^2 \,
\frac{
  \Gamma_\gamma^{\Sigma J_{\rm{IMS}} \rightarrow J_j \rightarrow J_{2^+}} \,
  \Gamma_\gamma^{\Sigma J_{\rm{IMS}} \rightarrow J_k \rightarrow J_{7^+}} \,
}
{\Gamma}
\label{eq:isigmastar}
\end{equation}
with the partial radiation width $\Gamma_\gamma^{\Sigma J_{\rm{IMS}}
  \rightarrow J_j \rightarrow J_{2^+}}$ summed over all partial widths
$\Gamma_\gamma^{J_{\rm{IMS}} \rightarrow J_j}$ leading finally to the $2^+$
isomer and a similar definition for $\Gamma_\gamma^{\Sigma J_{\rm{IMS}}
  \rightarrow J_k \rightarrow J_{7^+}}$ for transitions to the $7^+$ ground
state. Note that the properties of the thermally excited states with $J_j$ and
$J_k$ do not appear in Eq.~(\ref{eq:isigmastar}). The reason for this
cancellation is discussed in detail in \cite{Mohr07}. The spin factor in
Eq.~(\ref{eq:isigmastar}) is given by $g =
(2J_{\rm{IMS}}+1)/(2J_{\rm{g.s.}}+1) = 9/15 = 0.6$ \cite{Mohr07,Ward80}. For
the transition from the $7^+$ ground state to the $2^+$ isomer the integrated
cross section via the $4^+$ IMS is $I_\sigma^\ast = 133$\,eV\,fm$^2$ with an
uncertainty of about 20\,\% which is given by the 16\,\% uncertainty of the
lifetime of the IMS and the 13\,\% uncertainty of the branching ratio. Values
of the same order of magnitude have been obtained for integrated cross
sections in the neighboring odd-odd nucleus $^{108}$Ag \cite{Set16}, whereas
this result is about a factor of $20-100$ larger than the lowest
experimentally known IMSs in the heavy odd-odd nuclei $^{176}$Lu \cite{Mohr09}
and $^{180}$Ta \cite{Bel99,Bel02} which are located at slightly higher
excitation energies. The lowest IMS candidate in $^{180}$Ta is located at a
similar excitation energy as the $4^+$ IMS in \nbii ; its integrated cross
section is almost six orders of magnitude smaller \cite{Mohr07}. Also
relatively small values have been estimated for a low-lying $K$-mixing IMS in
$^{176}$Lu \cite{Gin09}. Very recently, IMSs have also been searched for in
$^{186}$Re \cite{Mat15}.

\section{Stellar reaction rate and effective half-life}
\label{sec:eff}
\begin{figure}[htb]
\includegraphics[width=0.90\columnwidth,bbllx=30,bblly=25,bburx=345,bbury=315,clip=]{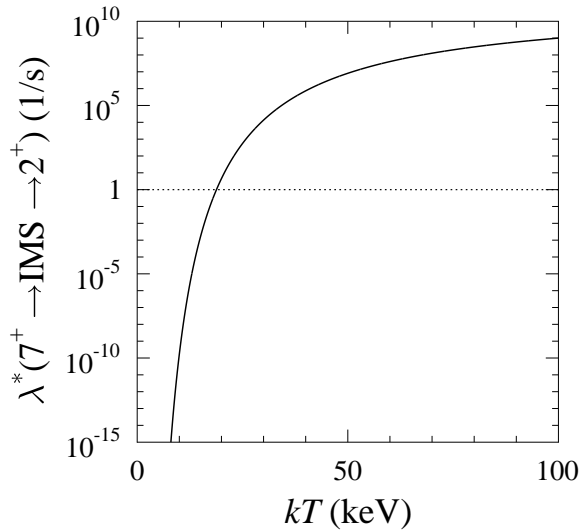}
\caption{
\label{fig:rate}
Stellar transition rate $\lambda^\ast$ for the transition from the $7^+$
ground state to the $2^+$ isomer via the $4^+$ IMS in \nbii . The horizontal
line at $\lambda^\ast = 1.0/s$ marks the typical timescale of supernova
explosions.
}
\end{figure}

The reaction rate $\lambda^\ast(T)$ for the $7^+ \rightarrow 2^+$ transition
via the $4^+$ IMS at $E^\ast = 480.3$\,keV is given by
\begin{eqnarray}
\lambda^\ast(T) & = & c \int n_\gamma(E_\gamma,T) \,
\sigma^\ast(E_\gamma) \, dE_\gamma \nonumber \\
& \approx & c \, n_\gamma(E^\ast,T) \, I_\sigma^\ast
\label{eq:rate}
\end{eqnarray}
with the integrated cross section $I^\ast_\sigma$ from
Eq.~(\ref{eq:isigmastar}) and thermal photon density
\begin{equation}
n_\gamma \, (E_\gamma,T) \, dE_\gamma = 
	\frac{1}{\pi^2} \, \frac{1}{(\hbar c)^3} \, 
	\frac{E_\gamma^2}{\exp{(E_\gamma/kT)} - 1} \, dE_\gamma .
\label{eq:planck}
\end{equation}
The result is shown in Fig.~\ref{fig:rate}. One finds a dramatic variation of
the transition rate as a function of temperature (given as thermal energy $kT$
throughout this paper). Already at $kT = 50$\,keV the rate exceeds $10^6$/s,
thus leading to thermalization of isomer and ground state under any typical
conditions of explosive nucleosynthesis. Consequently, the production ratio
between $2^+$ isomer and $7^+$ ground state of \nbii\ does not play any role
for the nucleosynthesis of \nbii\ because the properties of the $4^+$ IMS
ensure thermal equilibrium within less than 1 microsecond for any temperature
above $kT = 50$\,keV. In retrospect, this validates all previous studies which
did not take into account the isomer in the production of \nbii .

For completeness it should be mentioned that the rate for the reverse
transition from the $2^+$ isomer to the $7^+$ ground state can be
derived in the same way according to Eqs.~(\ref{eq:isigmastar}) and
(\ref{eq:planck}). As a result, the reverse rate is related to the forward
rate by detailed balance \cite{Ward80}.

After the explosive nucleosynthesis event the temperature drops.  As soon as
the rate $\lambda^\ast$ falls below the timescale of the explosive
nucleosynthesis event, thermal equilibrium between the $2^+$ isomer and $7^+$
ground state cannot persist. This happens at the relatively low temperature of
$kT \approx $ 19\,keV for a typical supernova explosion with timescales of the
order of one second. At this low temperature the ratio between isomer and
ground state is given by the Boltzmann factor $n(2^+)/n(7^+) = (5/15) \times
\exp{(-E^\ast/kT)} \lesssim 10^{-4}$. Thus, practically all \nbii\ survives in
the $7^+$ ground state. Because of the steep temperature dependence of the
reaction rate $\lambda^\ast$, the freeze-out temperature does not vary
dramatically for a broader range of explosion timescales. An increase
(decrease) of the explosive timescale by a factor of 10 reduces (increases)
the freeze-out temperature by less than 2\,keV which does not affect the
almost 100\,\% survival probability of \nbii\ in its $7^+$ ground state.

\begin{figure}[htb]
\includegraphics[width=0.90\columnwidth,bbllx=30,bblly=25,bburx=345,bbury=400,clip=]{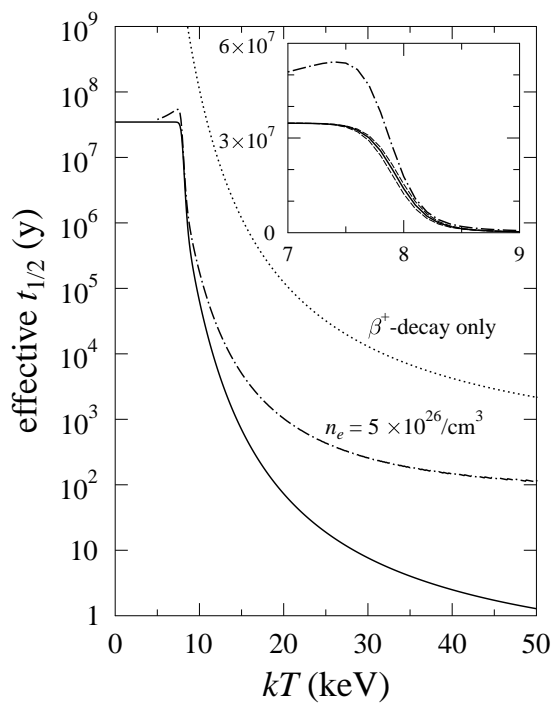}
\caption{
\label{fig:t12eff}
Effective half-life $t_{1/2}^{\rm{eff}}$ of \nbii\ as a function of
temperature. The inset shows the dramatic decrease of the effective half-life
around $kT \approx 8$\,keV and the nuclear uncertainties of a 20\,\% increased
or decreased coupling (dashed lines). An upper limit from
$\beta^+$-decay (without electron capture) is shown as dotted line. The
dash-dotted line represents an intermediate plasma density of $n_e = 5 \times
10^{26}$/cm$^3$. Further discussion see text.
}
\end{figure}

Although thermal equilibrium between the $7^+$ ground state and the $2^+$
isomer cannot be maintained at low temperatures, there is still a tiny
probability for a transition between ground state and isomer. This leads to a
temperature-dependent effective half-life of \nbii\ and may affect the
survival of \nbii\ in the cooling phase after its explosive production or in
any later re-heating. Qualitatively, as long as the transition rate
$\lambda^\ast$ is much smaller than the $\beta$-decay constant
$\lambda_\beta(7^+) = 6.3 \times 10^{-16}$/s, the effective half-life remains
constant at its laboratory value. At about 8\,keV $\lambda^\ast$ becomes
comparable to $\lambda_\beta(7^+)$, and thus the $2^+$ isomer is weakly
populated. Now the $\beta$-decay rate $\lambda_\beta(2^+) = 7.9 \times 10^{-7}$/s
is faster than the transition back to the ground state which leads to
$\beta$-decay of the isomer and a reduction of the effective half-life of
\nbii\ which scales with the transition rate $\lambda^\ast$. At higher
temperatures above 20\,keV, thermal equilibrium is established, and here the
effective half-life scales with the Boltzmann factor of the isomer and its
$\beta$-decay constant $\lambda_\beta(2^+)$. The resulting effective half-life
of \nbii\ is shown in Fig.~\ref{fig:t12eff}.

Unfortunately, there is a further complication. The $\beta$-decay constants
$\lambda_\beta(7^+$) and $\lambda_\beta(2^+)$ in \nbii\ are dominated by
electron capture (mainly from the $K$-shell). The binding energy of a
$K$-electron in \nbii\ is 19.0\,keV. Thus, at higher temperatures the
$\beta$-decay constants of ground state and isomer will decrease
significantly. This effect depends on the ionization state of \nbii\ which in
turn depends on temperature and electron density of the surrounding
plasma. Independent of temperature and density, the $2^+$ isomer has a small
$\beta^+$-decay branching with a branching ratio of $5.9 \times 10^{-4}$,
whereas $\beta^+$-decay has not been observed for the $7^+$ ground state. An
extreme upper limit of the effective half-life of \nbii\ can thus be
calculated from the $\beta^+$-decay branch of the isomer. This is shown as
dotted line in Fig.~\ref{fig:t12eff}.

It is difficult to provide a general estimate of the effective half-life of
\nbii\ under realistic astrophysical conditions as long as the astrophysical
scenario and the resulting plasma density are not known. Nevertheless, it can
be concluded that the steep drop of the effective half-life around $kT \approx
8$\,keV is real because at these temperatures the $K$-shell of \nbii\ is not
yet fully ionized. As an example, the effective half-life is also shown for an
intermediate electron density $n_e = 5 \times 10^{26}$/cm$^3$ in
Fig.~\ref{fig:t12eff}. One can see a small increase of the effective half-life
around $kT \approx 7$\,keV because of the starting ionization of the
$K$-shell. This increase is followed by a steep decrease around $kT \approx
8$\,keV which is governed by the transition rate from the $7^+$ ground state
to the $2^+$ isomer; even when partly ionized, the $\beta$-decay of the isomer
will still be sufficiently fast (see the above qualitative discussion of the
effective half-life).

Finally, the inset of Fig.~\ref{fig:t12eff} shows clearly that the dominating
uncertainties for the effective half-life of \nbii\ result from the
astrophysical conditions whereas the nuclear uncertainties are
marginal. Nevertheless, an independent confirmation of the unpublished
lifetime of the $4^+$ IMS in \nbii\ from \cite{USIP} is desirable.

\section{Conclusions}
\label{sec:conc}
The present study has shown that the low-lying $2^+$ isomer at $E^\ast =
135.5$\,keV does not play a significant role in the production of \nbii\ in
any explosive scenario. Thermal equilibrium is maintained down to low
temperatures below $kT \approx 19$\,keV where the Boltzmann population of the
$2^+$ isomer is already practically negligble. Thus the present study
validates in retrospect the previous nucleosynthesis studies of \nbii\ where
the isomer was not taken into account. However, the coupling to the isomer
leads to a dramatically reduced effective half-life of \nbii\ at temperatures
as low as $kT \approx 8$\,keV. Thus, any re-heating of the freshly synthesized
\nbii\ above 8\,keV, e.g.\ in the X-ray emitting supernova ejecta with its
typical temperatures in the low keV range \cite{Vink12,Bad10}, will reduce its
abundance significantly. This turns \nbii\ at least partly from a chronometer
to a thermometer for the thermal evolution after the last nucleosynthesis
event before the formation of our solar system.

\acknowledgments 
I thank U.\ Ott, M.\ Pignatari, Gy.\ Gy\"urky, and Zs.\ F\"ul\"op for
encouraging discussions. This work was supported by OTKA (K101328 and
K108459).

\end{document}